\documentclass[a4paper,fleqn,usenatbib]{mnras}

\usepackage[T1]{fontenc}
\usepackage{ae,aecompl}
\usepackage{graphicx}	
\usepackage{amsmath}	
\usepackage{amssymb}	
\title[Peculiar motion from quasar redshift distribution]{Peculiar motion of the solar system derived from a dipole anisotropy in the redshift distribution of distant quasars
}
\author[A. K. Singal]{Ashok K. Singal\thanks{E-mail: ashokkumar.singal@gmail.com}\\
{Astronomy and Astrophysics Division, Physical Research Laboratory, 
Navrangpura, Ahmedabad - 380009, India}}
\date{Accepted XXX. Received YYY; in original form ZZZ}
\pubyear{2019}
\begin{document}
\label{firstpage}
\pagerange{\pageref{firstpage}--\pageref{lastpage}}
\maketitle

\begin{abstract}
An observer stationary with respect to comoving coordinates of the expanding universe 
should find the redshift distribution to be isotropic. 
However, a peculiar motion of the observer would introduce a dipole anisotropy in the observed redshift distribution.   
Conversely, a dipole anisotropy in observed redshift distribution could be 
exploited to infer our peculiar motion, or rather of our solar system. 
We determine here our peculiar velocity by studying the dipole anisotropy 
in the redshift distribution of a large sample of quasars. 
The magnitude of the peculiar velocity thus determined turns out to be  
$2350\pm280$ km s$^{-1}$, not only much larger than 370 km s$^{-1}$ determined from the dipole anisotropy 
in the Cosmic Microwave Background Radiation (CMBR), but also nearly in an opposite direction. 
Such large values for peculiar velocity have been found in a couple of radio surveys too, 
but with a direction along the CMBR dipole. 
Large genuine differences in the inferred motion, whether in magnitude or direction, are rather disconcerting since 
a solar peculiar velocity should not depend upon the method of its determination. Such discordant dipoles 
imply perhaps an anisotropic universe, violating the cosmological principle, a cornerstone of the modern cosmology.
\end{abstract} 
\section{INTRODUCTION}
The peculiar velocity of the solar system relative to the frame of reference provided by the Cosmic Microwave 
Background Radiation (CMBR) was determined by COBE, WMAP and Planck satellites (Lineweaver et al. 1996; Hinshaw et al. 2009; Akrami et al. 2018) 
to be 370 km s$^{-1}$ in the direction RA$=168^{\circ}$, Dec$=-7^{\circ}$. 
However, from the sky brightness and number counts arising from discrete radio sources in the NRAO VLA Sky Survey (NVSS) dataset (Condon et al. 1998) comprising 1.8 million sources at 1400 MHz,  Singal (2011) found the solar motion to be $1600\pm400$ km s$^{-1}$, about 4 times larger than the CMBR value, though the direction (RA$=157^{\circ} \pm 9^{\circ}$, Dec$=-3^{\circ} \pm 8^{\circ}$) turned out within errors to be in agreement with the CMBR value. 
Subsequently many independent groups (Gibelyou \& Huterer 2012; Rubart \& Schwarz 2013; Tiwari et al. 2014; Colin et al. 2017; Bengaly, Maartens \& Santos 2018) have critically re-examined the two conflicting results and they also have found 
speeds $\sim 3-5$ times higher than the CMBR value, corroborating the claims of Singal (2011). This large difference in 
the CMBR and NVSS speeds is rather puzzling, more so as the direction of the motion turns out to be approximately the same in both cases. More recently, a similar investigation of angular distributions of number counts and the sky brightness of the TIFR GMRT Sky Survey (TGSS) - First Alternative Data Release (ADR1) dataset, comprising 0.62 million sources at 150 MHz (Intema et al 2017), 
have yielded the value of the peculiar velocity of the solar system to be an order of magnitude higher than the CMBR value (Bengaly et al. 2018; Singal 2019), though the direction again is coincident with that of the CMBR dipole.
Due to incongruous magnitudes of these dipoles, it is important if an independent confirmation of these results could be made. 

In order to resolve this germane issue of three conflicting magnitudes of solar motion inferred from different datasets using different techniques, 
we have attempted to determine here in an independent and rather more direct method, our peculiar motion from a dipole anisotropy, 
if any, in the redshift distribution of a large sample of quasars on the sky.
A distinct advantage with the redshift dipole is that it can provide a direct estimate of the solar peculiar velocity with respect to the reference frame of the 
quasars. It also obviates the need for looking for structure formations on large scales that might otherwise masquerade, for instance, as a number count dipole asymmetry purportedly due to observer's motion, in other indirect methods (see e.g., Rubart, Bacon \& Schwarz 2014; Tiwari \& Nusser 2016; Rameez et al. 2018).

\section{THE QUASAR SAMPLE}
We have derived our sample of quasars out of the DR12Q catalogue (P\^aris et al. 2017) from the Baryon Oscillation Spectroscopic Survey (BOSS) of the Sloan Digital Sky Survey (SDSS) III. 
The original catalogue contains 297301 quasars, with robust identifications and redshifts, measured by a combination of principal component eigenspectra, along with a visual inspection of their spectra. Out of these, a subset of 103245 quasars, listed with a uniformity index 1, implying that these form a homogeneously selected sample of quasars, form our present sample of quasars. For each of these objects, the sky positions (RA, Dec) and redshifts (z) from the BOSS pipeline with quoted errors, are available from the DR12Q FITS table file, downloadable from the SDSS public website (P\^aris et al. 2017).
Although a later version, DR14Q, has now become available with a much larger dataset of quasars (P\^aris et al. 2018), yet for our investigations DR12Q is still more appropriate because it includes the uniformity index that allows us to pick a homogeneously selected sample of quasars.
\section{Redshift dipole due to observer's motion}
The sky distribution of cosmological redshifts of a homogeneously selected sample of distant quasars should be statistically independent of the direction because of the assumed isotropy of the universe (\`{a} la cosmological principle).
However, an observer moving with a peculiar velocity $v$, defined as a motion relative to the comoving coordinates, will find a quasar lying at an angle $\theta$ with respect to the direction of motion, as seen in observer's frame, to have a redshift
\begin{equation}
(1+z)=(1+z_{\rm o})/\delta\approx (1+z_{\rm o})(1-(v/c)\cos\theta) 
\end{equation}
where $z$ is the measured redshift of the quasar, $z_{\rm o}$ is the cosmological redshift of the quasar with respect to the comoving rest frame at observer's location,
and $\delta=[\gamma(1-(v/c)\cos\theta]^{-1}$ is the Doppler factor due to the observer's 
motion with $\gamma=1/\sqrt{(1-(v/c)^2}$ as the Lorentz factor and $c$ as the 
speed of light in vacuum. In Eq. (1) we have used the non-relativistic formula for the Doppler factor since the CMBR, NVSS and TGSS results indicated that $v\ll c$. The observed redshift will then posses a dipole asymmetry
\begin{equation}
\Delta z= z-z_{\rm o}={\cal D}\cos\theta,
\end{equation}
with the dipole component ${\cal D}=-(1+z_{\rm o})(v/c)$, arising from the peculiar velocity $v$  of the observer (or equivalently of the solar system) with respect to the comoving rest frame. Since from the cosmological principle, the  
cosmological redshift $z_{\rm o}$ should have an isotropic distribution, it implies that if an average of the observed redshift $z$ 
is taken over different directions in the sky, then $\bar{z}={z}_{\rm o}$, with the dipole component getting cancelled. 

Since DR12Q sample does not have a uniform sky coverage (P\^aris et al. 2017), therefore, unlike in the case of NVSS or TGSS (Singal 2011; Gibelyou \& Huterer 2012; Rubart \& Schwarz 2013; Tiwari et al. 2014; Colin et al. 2017; Bengaly et al. 2018; Singal 2019), the 
direction of the dipole cannot be computed directly.
However, if we already have an idea of the the sky position ($\alpha_{\rm d},\delta_{\rm d}$) of the redshift dipole 
from some other considerations, viz. the dipole direction seen in the CMBR, NVSS or TGSS samples, then we can calculate the magnitude of the redshift dipole ${\cal D}$ and thereby the solar peculiar velocity $v$, and compare the value with those derived from these previous samples. The redshift distribution of distant quasars may thus provide an independent handle to hopefully resolve this pertinent issue of conflicting magnitudes of solar motion inferred from different datasets.

The procedure to be followed is straightforward. For every quasar position in the sample, we first calculate the polar angle $\theta$ with respect to the presumed dipole direction. Then a plot of the observed redshifts, $z$, against $\cos\theta$ should yield a linear correlation, $z=a+b\cos\theta$. Therefore, making a least square fit of a straight line to the ($\cos\theta,z$) data on quasars in our sample we can get ${z}_{\rm o}$ and ${\cal D}$ (with $z_{\rm o}=a$ and ${\cal D}=b$), which yields the magnitude of the solar peculiar velocity $v$ as
\begin{equation}
\frac{v}{c}=\frac{-{\cal D}}{1+z_{\rm o}}=\frac{-b}{1+a}.
\end{equation}
In fact, even if we have no a priori knowledge of ($\alpha_{\rm d},\delta_{\rm d}$), 
the sky position of the actual redshift dipole, for any arbitrarily chosen direction in sky, we can calculate this way, at least the projection of the actual redshift dipole in that particular direction, along with an estimate of the uncertainty in the dipole projection, from the least square fit procedure (Bevington \& Robinson 2003).


\begin{table*}
\begin{center}
\caption{Velocity components derived from the mean redshifts for various polar-angle bins of  $20^{\circ}$ width, about the CMBR dipole direction, RA$=168^{\circ}$, Dec$=-7^{\circ}$.}
\hskip4pc\vbox{\columnwidth=30pc
\footnotesize 
\begin{tabular}{ccccccccc}
\hline 
 Polar angle $\theta_{\rm i}$  & $N_{\rm i}$  & Mean projection    & Mean redshift  &  Redshift dipole component   & Velocity component along $\theta_{\rm i}$ \\
 range &   &  $\overline{\cos\theta}_{\rm i}$ & $\bar{z}_{\rm i}$ & $\Delta z_{\rm i}=\bar{z}_{\rm i}-{z}_{\rm o}$& 
$-c\Delta z_{\rm i}/(1+{z}_{\rm o})$ $(10^{3}$ km s$^{-1}$) \\ 
(1)&\,(2)&(3)&(4)&(5)&(6)\\ \hline
$0^\circ<\theta_1\leq 20^\circ$ & 3461 & 0.96 & 2.383 & $0.024\pm 0.011$ &  $-2.1\pm1.0$ \\
$20^\circ<\theta_2\leq 40^\circ$ & 19107 & 0.85 & 2.389 & $0.030\pm 0.005$  & $-2.7\pm0.4$ \\
$40^\circ<\theta_3\leq 60^\circ$ & 28720 & 0.64 & 2.374 & $0.015\pm 0.004$  & $-1.4\pm0.3$ \\
$60^\circ<\theta_4\leq 80^\circ$ & 20957 & $ 0.36 $ & 2.364 & $0.005\pm 0.005$ & $-0.4\pm0.4$ \\
$80^\circ<\theta_5\leq 100^\circ$ & 7176 & $0.04$ & 2.345 & $-0.014\pm 0.008$  & $1.3\pm0.7$ \\
$100^\circ<\theta_6\leq 120^\circ$ & 0 &  -- & -- &  --  & -- \\
$120^\circ<\theta_7\leq 140^\circ$ & 3680 & $-0.68$ & 2.345 & $-0.014\pm 0.011$ & $1.3\pm1.0$ \\
$140^\circ<\theta_8\leq 160^\circ$ & 11312 & $-0.87$ & 2.345 & $-0.014\pm 0.006$ &$1.2\pm0.6$ \\
$160^\circ<\theta_9\leq 180^\circ$ & 8832 & $-0.97$ & 2.331 & $-0.028\pm 0.007$ & $2.5\pm0.6$ \\
\hline
\end{tabular}
}
\end{center}
\end{table*}

\section{REDSHIFT DIPOLE MAGNITUDE ALONG THE DIRECTION OF CMBR DIPOLE}
Assuming, $\alpha_{\rm d}=168^{\circ}, \delta_{\rm d}=-7^{\circ}$, our sample yielded an average redshift ${z}_{\rm o}=\bar{z}=2.359 \pm 0.002$ and a finite value for the redshift dipole, ${\cal D}=0.026 \pm 0.003$, implying an excess of redshifts in the direction of the CMBR dipole.
Presumably ${\cal D}$ represents peculiar solar motion, accordingly from Eq. (3), we found a velocity component, $v=-2350 \pm 280$ km s$^{-1}$, in the direction of the CMBR dipole. 
In order to ensure that this dipole is not merely due to a skew distribution with a small number of stray extreme values, we divided 
the sky into  slices of $20^\circ$ wide bands in polar angle $\theta$ and for each $i^{th}$ band determined the mean redshift $\bar{z}_{\rm i}$ 
and its rms due to the spread in redshift within the band. The errors in redshifts of individual quasars are rather small ($\sim 0.003$, P\^aris et al. 2017) as compared to the spread of redshift in each $\theta$ bin. 
The variance in the redshift distribution about the mean value $\bar{z}_{\rm i}$ among $N_{\rm i}$ quasars for each $i^{th}$ bin 
is $\sigma_{\rm i}^2=\Sigma(z_{\rm j}- \bar{z}_{\rm i})^2/(N_{\rm i}-1)$, 
where the summations are over $j$ that runs from 1 to $N_{\rm i}$ quasars for each $i^{th}$ bin, and the standard deviation of the mean ${\Delta z}_{\rm i}$  then is $\sigma_{\rm i}/\sqrt{N_{\rm i}}$ (Bevington \& Robinson 2003).

\begin{figure}
\includegraphics[width=\columnwidth]{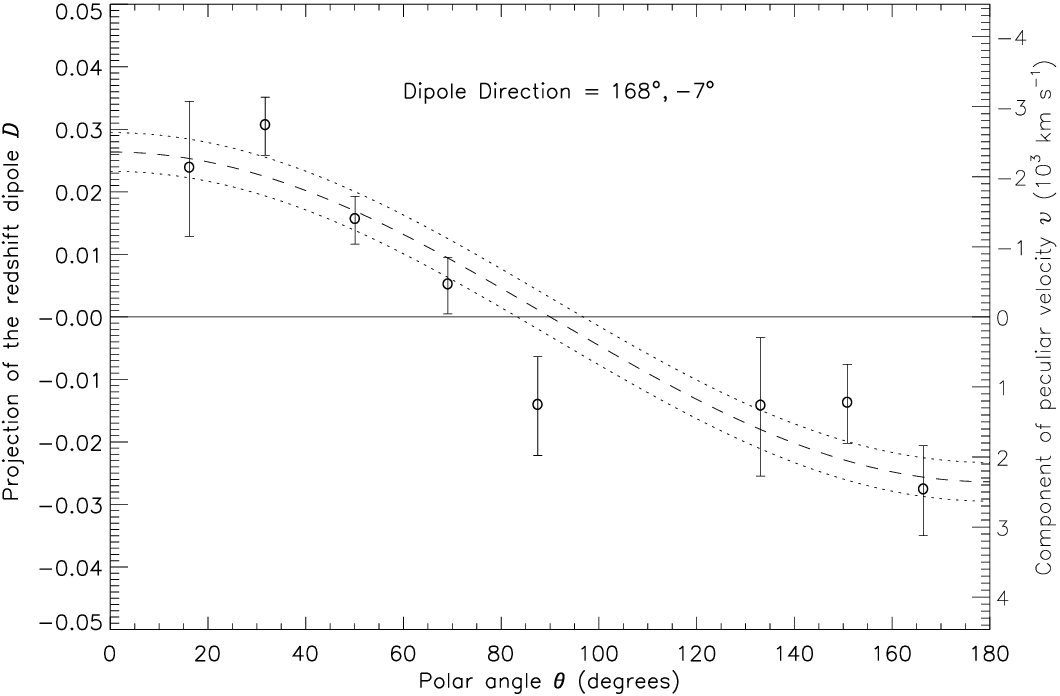}
\caption{A plot of the observed redshift dipole components for various polar angles, computed with respect to the direction (RA$=168^{\circ}$, Dec$=-7^{\circ}$) of the CMBR dipole. Circles (o) with error bars represent
values for bin averages of the redshift dipole components, obtained for various $20^{\circ}$ wide slices of the sky. 
The corresponding peculiar velocity of the solar system is shown on the right hand scale. The dashed line 
shows the least square fit to the whole sample of 103245 quasars. The two dotted lines show the $1 \sigma$ spread around the least square fit.}
\end{figure}

The computed quantities during each step for various $\theta$ bins are presented in Table~1, organized in the following manner. 
(1) Polar angle range for the $i^{th}$ bin.
(2) Number of quasars $N_{\rm i}$ in the $i^{th}$ bin.
(3) Mean projection factor over all $N_{\rm i}$ sources in the bin, $\overline{\cos \theta}_{\rm i}=\Sigma{\cos\theta}_{\rm j}/N_{\rm i}$.
(4) Mean value of redshift for the $i^{th}$ bin.
(5) Mean value of the component of the redshift dipole, $\Delta z_{\rm i}=\bar{z}_{\rm i}-{z}_{\rm o}$, obtained by subtracting the  mean value of the redshift  
${z}_{\rm o}=\bar{z}=2.359$, computed for the whole sample of 103245 quasars, from the average redshift, $\bar{z}_{\rm i}$ for the bin in column (4).
(6) Component of the peculiar solar velocity component $v_{\rm i}\approx v\,\overline{\cos \theta}_{\rm i}$, in units of ($10^{3} $ km s$^{-1}$), along the direction of the $i^{th}$ bin, obtained by multiplying the redshift dipole component ${\Delta z_{\rm i}}$ in column (5) by $(-c/1+{z}_{\rm o})$. The negative sign indicates that the inferred direction of the peculiar motion of the observer is in a direction opposite to that of the redshift dipole ${\cal D}$ (cf. Eq. (3)).

These values are plotted in Fig.~1, and as can be seen from the figure, there is a systematic  variation, $\propto \cos\theta$, in the estimated 
solar motion component for successive bins, matching with the least square fit to the ($\cos\theta,z$) data for all 103245 quasars shown as a dashed line. 
We get a peculiar velocity component of magnitude $-2350 \pm 280$ km s$^{-1}$ in the assumed direction (RA$=168^{\circ}$, Dec$=-7^{\circ}$) of the redshift dipole. This furnishes a first direct evidence of the presence of a dipole in the quasar redshift data, with a systematic excess of redshift in the direction of the CMBR dipole. Though the ultimate direction of the redshift dipole is not yet established, still an unexpectedly large, negative peculiar velocity component towards the CMBR dipole direction is quite intriguing.
\section{DIRECTION OF THE REDSHIFT DIPOLE}
To ascertain the real direction of the redshift dipole, which may actually lie elsewhere in the sky, 
we divided the sky into a grid of $10^\circ \times 10^\circ$ bin size with the total 41253 square degrees ($=4 \pi$ sr) 
area of sky covered with $422$ cells (with a slight overlap between RA$=360^{\circ}$ and $=0^{\circ}$). 
Then, with respect to the direction of each of these 422 grid points in turn, we calculated the polar angle $\theta$ for 
all 103245 quasar positions in the sky, and made a least square fit of a straight line to the ($\cos\theta,z$) data on 
these 103245 quasars to estimate ${\cal D}$, as described earlier. This actually yielded a projection of the redshift dipole 
${\cal D}$, along with its standard error, in the direction of each of our 422 grid points in sky. 

For a good sky coverage of quasar positions, this itself should suffice to arrive at the direction of the redshift dipole from the peak among 422 ${\cal D}$ values. However, owing to a non-uniform sky coverage with the quasars and the statistical spread in the redshift values, an unambiguous unique peak was not discernible. However, it was possible to constrain the search further by noting that with respect to an actual dipole direction ($\alpha_{\rm d}, \delta_{\rm d}$) there should be a $\cos\theta$ dependence in the 422 ${\cal D}$ values, where $\theta$ is the polar angle for the corresponding grid point, with respect to $\alpha_{\rm d}, \delta_{\rm d}$. 

To treat all directions at this stage on equal footing as prospective dipole directions, we tried in an unbiased manner `the brute force method'. Starting with RA$=0^{\circ}$, Dec$=0^{\circ}$ and progressing in steps of $0.5^{\circ}$ in both RA and Dec, for each of these direction we calculated $\theta$ for every grid point and made a least square fit of a straight line to the ($\cos\theta,{\cal D}$) data, weighted by the inverse of the variance of ${\cal D}$, of 422 grid points. The process converged quickly to show a clear unique peak, in the vicinity of the minimum chi-square value, the latter at RA$=169^{\circ}$, Dec$=-3^{\circ}$. The peak occurred at RA$=164^{\circ}$, Dec$=-21^{\circ}$, with a height, $3.6\times 10^{-2}$. Figure (2) shows the peak and the neighbourhood around it. In order to make sure, we also tried a finer grid with $2^\circ \times 2^\circ$ bins with $10360$ cells, but it made no perceptible difference in our results. 
Figure (3) shows the ($\cos\theta,{\cal D}$) scatter plot of 422 grid points and the least square fit to this data for the dipole along RA$=169^{\circ}$, Dec$=-3^{\circ}$, the direction with a minimum chi-square value. 
The fit to the data seems very good. Only in a very contrived situation could have we obtained such consistent results for the fit for various directions if the redshift dipole were not really present in our data.

\begin{figure}
\includegraphics[width=\columnwidth]{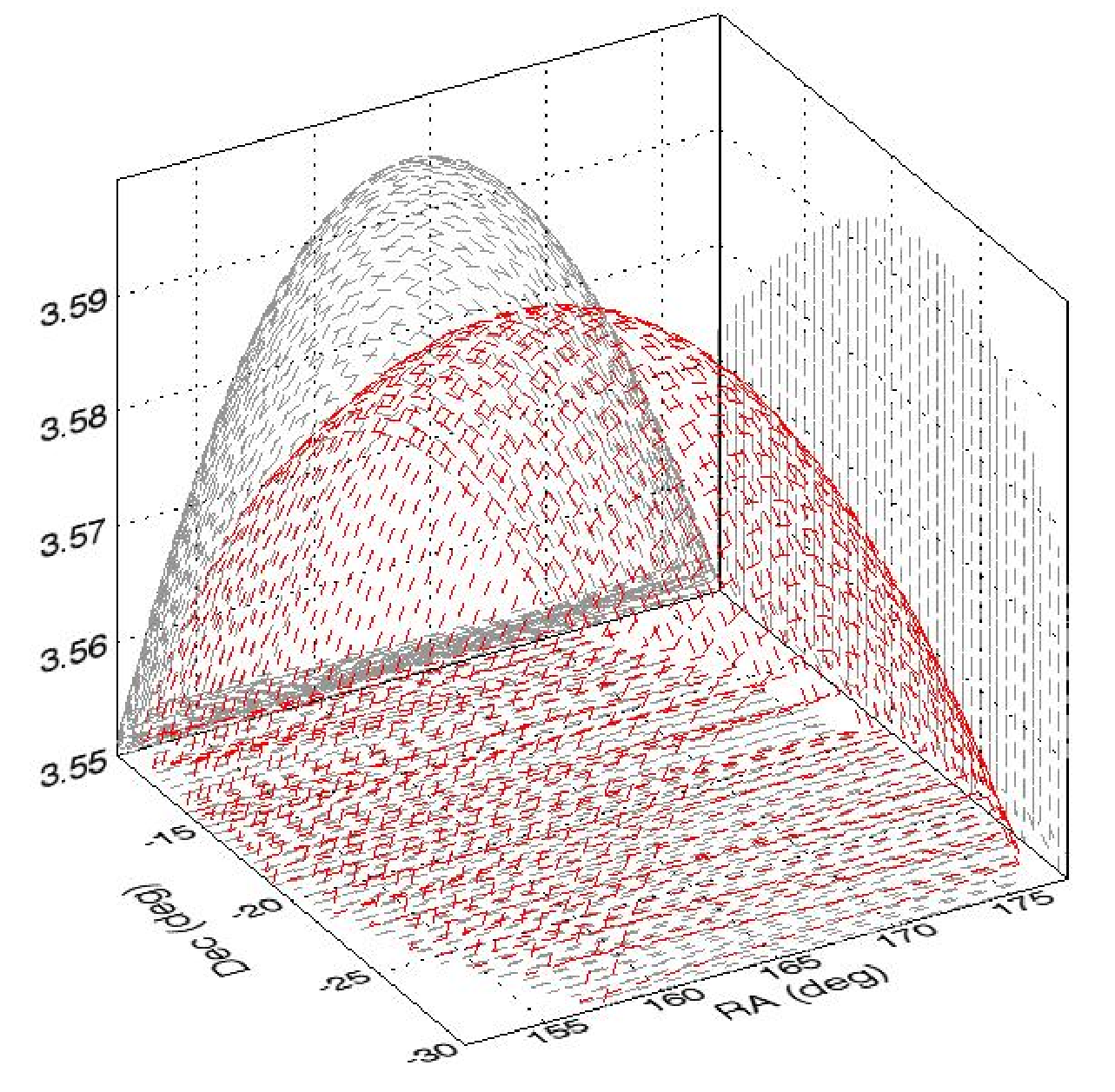}
\caption{A 3-d plot showing trials results (in red colour) of prospective dipole directions. RA and Dec are both expressed in degrees, while the vertical axis represents the redshift dipole magnitude ${\cal D}$ (in units of $10^{-2}$). The plot shows the drop ($\propto \cos\theta$) seen for neighbouring sky directions around a clear, single peak, which approximately coincides with the minimum chi-square value. The 2-d projections of the plot in RA-${\cal D}$ and Dec-${\cal D}$ planes, shown in light grey, identify coordinates of the peak at RA$=164^{\circ}$, and Dec$=-21^{\circ}$.} 
\end{figure}

It may be added that there are also 174831 quasars in the DR12Q catalogue, listed with a uniformity index 0, implying these are not part of a homogeneously selected sample. In fact, a trial with this non-homogeneous sample of quasars did not yield any redshift dipole with a statistical significant magnitude, as should indeed be expected. 

To test our procedure as well as to rule out any subtle unknown effects, if any, due to the incomplete sky coverage, we did Monte Carlo 
trials, for the given quasar positions (thereby keeping the same sky coverage) but with a random redistribution of the 
redshifts observed in the sample. One hundred different trials with a random redistribution of the redshifts were made, in no case any significant redshift dipole was found in any direction in the sky. Thereafter an artificial dipole due to a hypothetical solar motion of a randomly selected magnitude and direction,  
was superimposed on each of these 100 randomly redistributed redshift datasets, and our procedure was repeated to see if we do retrieve 
the values that were input in each realization. Not only did we recover the input values, both in direction and magnitude, thereby verifying our procedure as well as the computer routine, these simulations also allowed us to make an estimate of errors in the derived dipole direction. 

\begin{figure}
\includegraphics[width=\columnwidth]{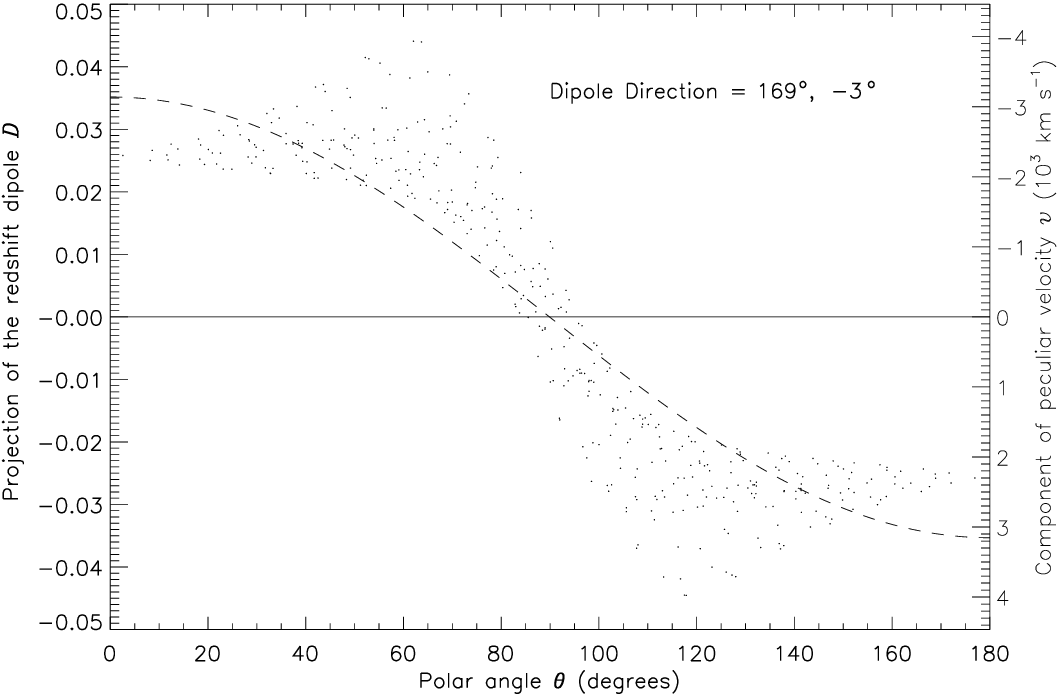}
\caption{A polar angle-redshift dipole projection ($\theta-\cal D$) scatter diagram of the 422 grid points, about the minimum Chi-square fit along the dipole direction 
RA$=169^{\circ}$, Dec$=-3^{\circ}$.}
\end{figure}
Accordingly, we arrive at $\alpha=164^{\circ}$, $\delta=-21^{\circ}$
from the peak in dipole magnitude, or $\alpha=169^{\circ}$, $\delta=-3^{\circ}$
from the minimum value of the Chi-square fit. Estimated errors in either case are $\Delta\alpha=\pm 10^{\circ}$ and $\Delta\delta= \pm15^{\circ}$. Both these values for the redshift dipole direction agree, within the $1\sigma$ uncertainty, with the CMBR dipole direction, RA$=168^{\circ}$, Dec$=-7^{\circ}$, which we adopt as $\alpha_{\rm d}, \delta_{\rm d}$ for the redshift dipole. It does seem quite convincing that the redshift dipole is genuine, otherwise, we could not have got this `out of the blue' appearance of the redshift dipole direction coinciding with the CMBR dipole direction, except perhaps in a very contrived situation. 

Magnitude of the redshift dipole from Fig. (1) corresponds to a solar speed $2350\pm 280$ km s$^{-1}$, an $8\sigma$ result with $\stackrel{>}{_{\sim}} 6$ times the CMBR value. However, the motion is in a direction opposite to that of the CMBR dipole, thereby implying a peculiar motion in sky along RA$=348^{\circ}$, Dec$=7^{\circ}$.  

\begin{table*}
\caption{Peculiar velocity, $v$, of the solar system in the CMBR dipole direction, derived using different methods from various datasets.}
\begin{tabular}{@{}cccccccc}
\hline
Dataset & Observation &  physical effect involved & $v$ (km s$^{-1}$)& Reference\\
(1)&(2)&(3)&(4)&(5)\\
\hline
CMBR & Dipole anisotropy in CMB temperature & Doppler effect  due to Observer's motion & 370 & Akrami et al. (2018)\\
NVSS & Dipole in source counts at 1400 MHz & Aberration and Doppler effect & 1430 & Singal (2011, 2019)\\
TGSS & Dipole in source counts at 150 MHz & Aberration and Doppler effect & 3770 & Singal (2019)\\
DR12Q & Dipole anisotropy in quasar redshifts & Redshift change due to Observer's motion  & -2350  & present work\\\\
\hline
\end{tabular}
\end{table*}
\section{DISCUSSION}
Like in the NVSS and TGSS cases (Singal 2011; Gibelyou \& Huterer 2012; Rubart \& Schwarz 2013; Tiwari et al. 2014; Colin et al. 2017; Bengaly et al. 2018; Singal 2019), here too the derived direction of the dipole seems to coincide, within errors, with that of the CMBR dipole, with a systematic excess of redshift in this direction. But there is an important difference in the redshift dipole case as the inferred peculiar velocity of the solar system is anti-parallel to all other earlier velocity estimates, including that of the CMBR. From these four dipoles we cannot arrive at a single coherent picture of the solar peculiar velocity, which,   
defined as a motion relative to the local comoving coordinates  and from the cosmological principle, a motion with respect to an average universe, should not depend upon the exact method used for its determination. Therefore such large discrepancies in the inferred velocity vectors may perhaps be a pointer to the need for some rethinking on the conventional interpretation of these dipoles. 

The results for various dipoles are summarized in Table 2, which is organized in the
following manner: 
(1) Dataset used.
(2) The observation done.
(3) The physical effect involved.
(4) Inferred peculiar velocity of the solar system.
(5) Reference to entries in columns (1) to (4).
From Table 2, we see that the peculiar velocity of the solar system, derived from different datasets, using different techniques,  not only differs vastly in magnitude, but it also seems to be in opposite direction, at least in one of the cases. 

The rms error for the peculiar velocity estimates for the NVSS, TGSS and DR12Q   
are $\sim 300-400$ km s$^{-1}$ (Singal 2011, 2019), and a positive detection of a velocity $\sim 370$ km s$^{-1}$, (similar to the CMBR value) would not have been possible in such surveys (Crawford 2009). A couple of earlier attempts had concluded the radio source dipoles to be consistent with the CMBR dipole as no unambiguous, statistical significant, departures from the CMBR dipole value could be established (Baleisis et al. 1998; Blake \& Wall 2002). 
However, a peculiar velocity $\sim 4$ times larger than the CMBR value was detected in the NVSS dataset at a statistically significant level (Singal 2011), soon to be confirmed by many independent groups (Gibelyou \& Huterer 2012; Rubart \& Schwarz 2013; Tiwari et al. 2014; Colin et al. 2017; Bengaly et al. 2018), and even a larger peculiar velocity was recently determined from the TGSS data (Bengaly et al. 2018; Singal 2019). These detections became possible only because the dipole strength turned out to be much larger than the CMBR value, so that the signal in each case was way above the rms error. Here, we had hoped to get an independent handle in the redshift distribution of distant quasars to be able to resolve this germane issue of conflicting magnitudes of solar motion inferred from different datasets. But the problem seems to have become all the more intricate now.

The observed fact that these four discordant dipoles, resulting from four different surveys carried out with independent instruments and techniques in different wavebands by different independent groups, happen to be pointing along the same direction in sky, shows that these dipoles are not results of some systematics in individual surveys (otherwise they could have been pointing in random directions in sky). Further, it also indicates that this particular direction in sky does have some peculiarity and in some respects is a preferred direction, though at this stage it might be premature to call it some sort of an ``axis'' of the universe. Nevertheless, four independent dipole vectors pointing along the same particular direction could imply an anisotropic universe, violating the cosmological principle, a cornerstone of the modern cosmology. 

\section{CONCLUSIONS}
A redshift anisotropy in a homogeneously selected sample of 103245 quasars, was detected, which showed a systematic excess of redshift towards the direction of the CMBR dipole. The redshift dipole is found, within errors, aligned with the CMBR dipole along RA$=168^{\circ}$, Dec$=-7^{\circ}$. 
The inferred peculiar velocity of the solar system, $\sim 6$ times larger than the CMBR value, is pointing, accordingly, in a direction opposite to that of the CMBR dipole. 
Earlier peculiar velocity vectors, determined from dipoles in radio source number counts in the NVSS and TGSS, were found  to be $\sim 4$ and $\sim 10$ times larger than the CMBR value, though pointing along the same direction as the CMBR dipole. 
Since a solar peculiar velocity should not depend upon the technique used for its determination, some alternate explanation for these discordant dipoles may be required. Nonetheless, the   emergence of the same preferred cosmic direction repeatedly in different dipoles, implies perhaps an anisotropic universe, violating the cosmological principle, a cornerstone of the modern cosmology.

\end{document}